# HIGGS BOSON MUON COLLIDER FACTORY: $h^0$, A, H STUDIES *

D. Cline[#], X. Ding, J. Lederman, UCLA, Los Angeles, CA 90095, USA

*Abstract*

With the recent hints of the Higgs boson from the LHC and a mass near 125 GeV/c we re-propose to study and build a muon collider Higgs factory to study the Higgs in the S channel [1]. This was first proposed in 1992 by the first author. It is essential to study the Higgs boson for clues to new physics. The formation of the DOE MAP program, recent advances in 6D μ cooling methods, simulation, and targeting make this a feasible project to initiate at this time. This collider would fit into the FNAL site.

## THE FIRST STUDIES OF MUON COLLIDER HIGGS FACTORY 1992-1994

The early studies of muon collider focussed on an S channel Higgs Factory ($h^0$) [2]. The major scientific goal is to measure the exact $h^0$ mass and the width in the S channel. There were several workshops devoted to the method to carry out on energy scan to find the very narrow Higgs Boson. We show several of the plots in the initial Higgs studies in Figure 1-2 and Table 1, which gives a strong argument for a Higgs muon collider [2].

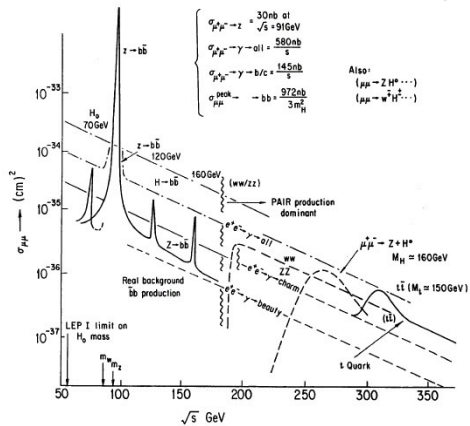

Figure 1. Muon Collider Higgs boson factory: $h^0$ Higgs boson $m_{h0}$ ~120 GeV.

Table 1. Arguments for a Higgs-Factory $\mu^+\mu^-$ collider

1. The $m_\mu / m_e$ ratio gives coupling 40,000 times greater to the Higgs particle. In the SUSY model, Higgs $m_h < 120$ GeV!!
2. The low radiation of the beams makes precision energy scans possible.
3. The cost of a "custom" collider ring is a small fraction of the $\mu^\pm$ source.
4. Feasibility report to Snowmass established that $\mathcal{L} \sim 10^{33}$ cm$^{-2}$ s$^{-1}$ is feasible.

With the current evidence for a Higgs Boson with a mass of 125 GeV from the CMS and ATLAS detectors [1] and the Tevatron makes it important to study the feasibility of a muon collider Higgs Factory. Most SUSY models predicted a Higgs mass of 120 GeV. Thus the evidence for a 125 GeV Higgs is a surprise.

We have devised a ring cooler for 6D cooling [2]. The key idea behind a muon collider Higgs factory is to be able to measure the exact mass and width in this channel. We showed in 1993-1994 that this is possible.

In Reference 4 we follow a discussion of the type of physics process that a muon collider exceeds at carrying out. As shown in Figures 1 and 2 the low mass $h_0$ Higgs is very narrow. Reference 4 suggests an energy scan strategy to find this narrow resonance with a muon collider.

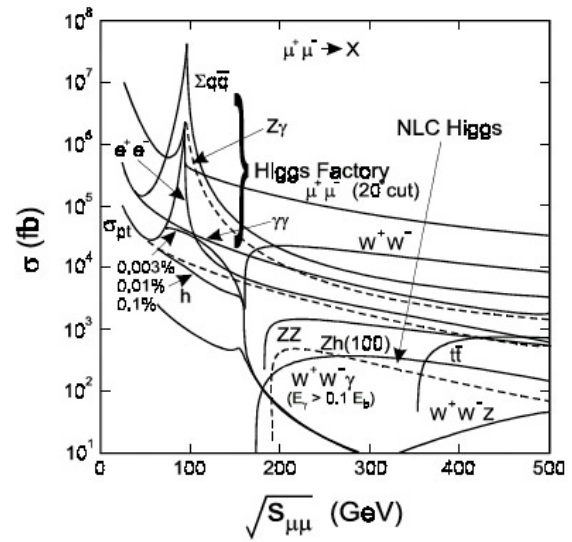

Figure 2. Comparison of $\mu^+\mu^-$ and $e^+e^-$ production.

## THE STUDY OF THE SUPERSYMMETRIC A AND H HIGGS BOSONS WITH A MUON COLLIDER: A/H HIGGS FACTORY

In the mid 1990's a study of the A/H Higgs Bosons was carried out. If the possible observation of a 125 GeV Higgs Boson implies the existence of Super Symmetry in Nature as many think; the mass of the S quarks could be rather large, possibly at the limit of observation at the current LHC. It is still possible that the A/H Higgs Bosons could be lower mass and observable. A muon collider could then study the A and H in the S channel. The A/H widths will be larger and easier to observe than $h^0$. See Figure 3 [4]. Polarized beams may help [5].

We have shown that the A and H could have a CP violating interaction that could be studied in the muon collider A/H Higgs Factory. The mass of the A/H may be much larger than shown in Figure 3. One guess is 800 – 1000 GeV [6].

We have also studied the possible detection of CP violation between the CP even and off A/H states. One could either study $\tau\bar{\tau}$ final states or use partially polarized μ in the muon collider (see Reference 5).

## CP VIOLATION

The use of polarized muons can lead to a test for CP violation in the interference between the A and H particles. This could be of key importance [6] to understanding the origins of CP violations.

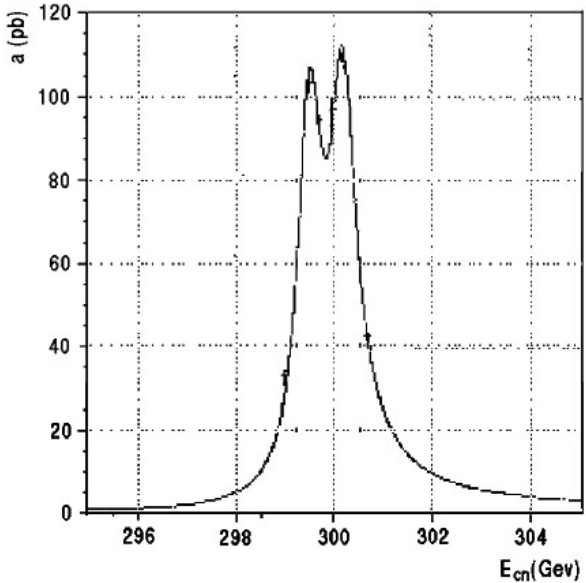

Figure 3. A/H Higgs boson factory to observe CP violation.

## CURRENT STATUS OF THE STUDIES OF A MUON COLLIDER

There has been a great deal of progress in the development of a muon collider. We show a schematic of the muon collider in Figure 4. The key issue is 6D cooling of the muons. Figure 5 shows the emittance that may be achieved by the various 6D cooling scheme. The UCLA/BNL etc. team has studied a Ring Cooler and shown with careful simulation that robust cooling could be achieved [3]-other schemes also show great promise.

While Figure 4 shows a 3 TeV muon collider a Higgs factory collider may be easier to make (125 GeV). The need to accelerate to TeV energies is commoner (only to 62.5) while the 6D cooling will be the same. Such a system could either be upgraded to an A/H factory (TeV) or high energy muon collider (multi TeV).

One example of 6D cooling can be found in Reference 3. We use a ring cooler made up of solenoids and dipoles. Robust 6D cooling is show in the simulation in Reference 3 (see Figure 7) [3].

Final cooling is also of key importance (Figure 5). J. Lederman is studying final cooling using high field solenoids (40T) shown in Figure 6 (see Reference 7).

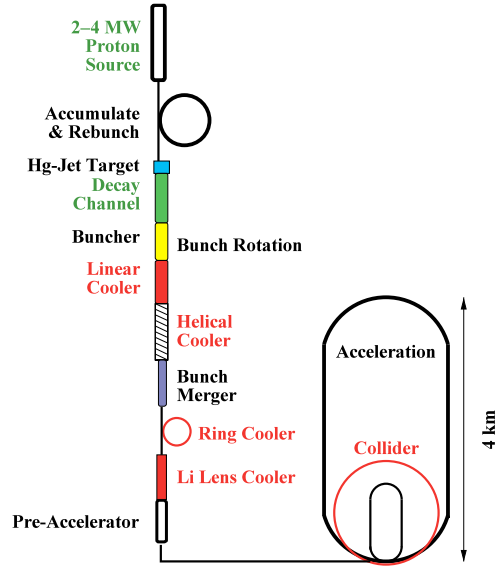

Figure 4. Schematic of a 1.5-TeV muon collider.

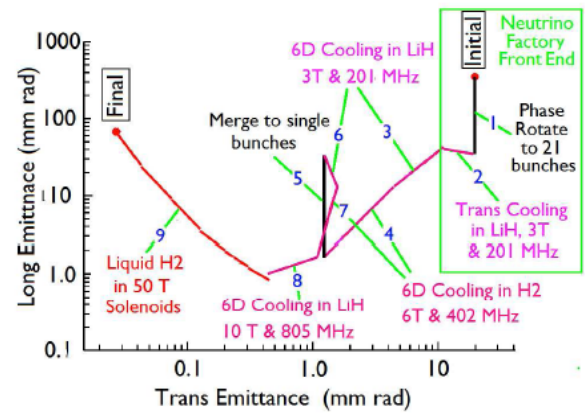

Figure 5. A complete scheme of ionization cooling for a muon collider.

## POSSIBLE 6D COOLING ADVANCES

We show a Ring Cooler for 6D cooling and a Final Cooler that uses a 40T Solenoids in Fig. 6 and 7.

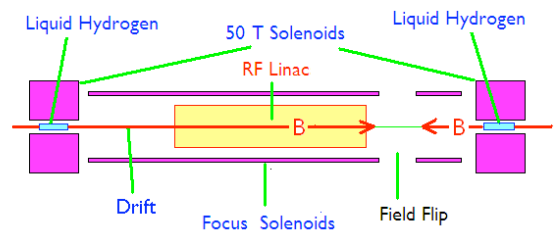

Figure 6. Final cooling scheme using 50T solenoids

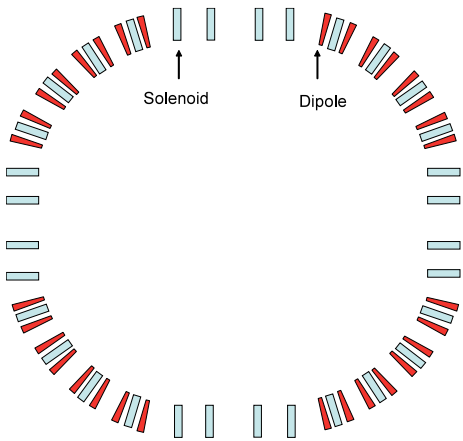

Figure 7. Schematic drawing of a four-sided ring using dipoles-solenoids for muon 6D cooling. The ring cooler 6D system is fully described in Reference 3.

## SUMMARY

Now that evidence is presented for a Higgs boson of mass 125 GeV it is time to study a muon collider Higgs factory. The U.S. DOE has started the MAP project along with other labs in the USA. The possible study of A/H supersymmetric Higgs bosons offers the possibility to detect CP violation in a new way.

## ACKNOWLEDGEMENT

We wish to thank Gail Hanson for many discussions on this topic.